\documentstyle[12pt,epsf]{article}

\date{July 26, 1996}
\newcommand{\tenrm}{\small}
\newcommand{\be}{\begin{equation}}
\newcommand{\ee}{\end{equation}}
\newcommand{\bea}{\begin{eqnarray}}
\newcommand{\eea}{\end{eqnarray}}

\begin{document}
\title{ 
QCD at finite temperature and partially negative flavour numbers
}

\author{G.\ M.\ de Divitiis, R.\ Frezzotti, M.\ Masetti and R.\ Petronzio \\
\small Dipartimento di Fisica, Universit\`a di Roma {\em Tor Vergata} \\
\small and \\
\small INFN, Sezione di Roma II \\
\small Viale della Ricerca Scientifica, 00133 Roma, Italy \\
\medskip
}

\maketitle

\begin{abstract} 
We study dynamical fermion effects in lattice QCD at finite temperature. 
The method adopted is basically the extrapolation from negative flavour 
numbers already tested at zero temperature and based on the simulation 
of local bosonic theories, with an essential
difference. With an appropriate choice of the boundary 
conditions on the bosonic fields, called ``bermions'', it is 
possible to separate the $Z_3$ breaking contribution of fermion loops to 
the effective action from the one conserving $Z_3$: the former
is simulated exactly at a fixed  positive and even flavour number, 
while the extrapolation from negative 
flavour numbers is made only on the $Z_3$ invariant part of the action. 
We test this approach by comparing our results on a $16^3\times 2$ lattice 
with those from a hopping parameter expansion
and our results on a $16^3\times 4$ lattice with those of direct 
Monte Carlo simulations including the fermion determinant.

 \end{abstract}

\vfill

\begin{flushright}
  {\bf ROM2F-96-41 ~~~~}\\
\end{flushright}

\newpage

\section{Introduction} 

A new approach to the problem of estimating dynamical fermion effects in 
lattice QCD based on the extrapolation from theories with a negative number 
of dynamical flavours was recently proposed and applied in numerical 
simulations of the theory at zero temperature. In that case the quenched 
approximation appears to reproduce most of the experimental pattern of the 
hadronic spectrum and unquenching corrections are small when the comparison 
with the quenched case is performed after a suitable shift of the bare 
parameters i.\ e.\ at fixed values of the lattice spacing and of the 
renormalized quark mass. The ``bermion'' method exploits the smoothness of 
dynamical flavour dependence studied at fixed renormalized quantities to make 
estimates in full QCD which are extrapolations of results obtained at negative 
flavour numbers, where fermions are replaced by ``bermions'', i.e. bosons with 
a fermion action \cite{LL, Noi}. 

In this paper we extend this approach to QCD at finite temperature. The 
unquenching effects are not expected in this case to reduce to a simple 
renormalization with some small residual corrections. 
The $SU(3)$ pure gauge 
theory is invariant under the discrete group $Z_3$ and the deconfining 
transition in the theory is associated with the spontaneous breaking of 
this symmetry. In full QCD the $Z_3$ symmetry is broken explicitely 
by fermion loops and the corresponding phase transition could disappear 
for intermediate values of quark masses. The straightforward application 
of the bermion method to finite temperature QCD would require an extrapolation 
from negative to positive flavour numbers $n_f$ also for the $Z_3$ breaking 
fermion loop contribution to the effective action. This extrapolation seems 
quite unsafe because thermodynamical QCD observables, e.\ g.\ the critical 
temperature, are not expected to behave smoothly around $n_f = 0$. However it 
is possible, using appropriate boundary conditions on the bermion fields, to 
simulate theories where the $Z_3$ breaking and the $Z_3$ conserving 
contributions of fermion loops to the effective action get different 
weights corresponding to different ``effective flavour numbers''. 
As a result, the $Z_3$ breaking interaction can be simulated exactly 
for even and positive flavour numbers, while the $Z_3$ conserving 
dynamical fermion contribution to the action, which is expected to 
give mainly a renormalization effect, is extrapolated from negative 
flavour numbers. 

In section 2 we present the bermion action and the choice of the boundary 
conditions which leads to the correct $Z_3$ breaking interaction.
 In section 3 
we compare our results on a $16^3\times 2$ lattice with those of a
hopping parameter expansion \cite{karsch} and our
results on a $16^3\times 4$ lattice with those of direct Monte Carlo 
simulations including the fermion determinant for $N_t = 4$ \cite{frabernardo}. 

\section{The role of the boundary conditions} 

The action for lattice QCD with $n_f$ flavours of Wilson fermion is 
\be S[U,\psi,\bar\psi] = S_G[U] + 
\sum_{j=1}^{n_f} \sum_x \bar\psi_j(x)\gamma_5 [Q\psi_j](x) \ee 
where $S_G[U]$ is the gauge action, $j$ is a flavour index and $\gamma_5Q$ 
is the Dirac operator: 
\bea
[Q\psi_i](x) = \frac{1}{2\kappa}\gamma_5 \psi_i(x)
- \frac{1}{2}\gamma_5\sum_{\mu=0}^3
U_\mu(x)(1-\gamma_\mu)\psi_i(x+\mu) \nonumber \\
- \frac{1}{2}\gamma_5\sum_{\mu=0}^3 
U^\dagger_\mu(x-\mu)(1+\gamma_\mu)\psi_i(x-\mu) 
\eea
After integration over the quark fields one obtains an effective action:
\be S_{eff}[U] = S_G[U] - n_f Tr ( \log \gamma_5 Q ) \ee 
where the second term is a sum of one loop diagrams.
These can be divided into two classes: to the first
belong the loops which lie entirely inside
the lattice or wrap around the edges of the lattice a
multiple of three times closing
through the boundary conditions and to the second those which wrap a number 
of times different from multiples of three.
At zero temperature all winding loops 
are just a finite volume effect, but at finite temperature, i.e. at
finite physical extent of the ``time'' direction, the loops winding in this 
direction and belonging to the second class do 
influence the existence of the deconfining phase transition and 
an extrapolation from negative flavour numbers cannot be expected to
be smooth. However, there are choices of the boundary
conditions which affect only the second class and which reproduce
the correct fermionic $Z_3$ breaking contribution also in the case where the 
fermion determinant is replaced by a bermion one.

The action for QCD with a negative flavour number $n_f$ is written in terms 
of $n_b = |n_f|/2$ bermions fields $\phi_j(x)$ which are commuting spinors:
\be S_b[U,\phi,\phi^\dagger] = S_G[U]+\sum_{x,y,z}\sum_{j=1}^{n_b} 
\phi_j^\dagger(x)Q(x,z)Q(z,y)\phi_j(y) \ee
where $j$ is a flavour index. 
We introduce a set of {\it flavour dependent} boundary conditions 
parametrized by the elements of the
center of the gauge group $Z_3$:
\be \phi_j({\mbox \bf x},t+N_t) = - z_j \phi_j({\mbox \bf x},t) ~~~~~~~~~~~
z_j \in Z_3\ee 
The contribution of $n_b$ flavours to the second class of bermion loops in 
the effective action can be expressed as a sum of terms which are 
characterized by winding number modulo three $i$ of Polyakov loop-like 
terms $P_i$ in the ``time'' direction. The full expression for $S_{eff}$ 
can be parametrized as:
\be S_{eff} = S_G +n_b Re[P_0] 
+ Re\Big[\Big(\sum_{j=1}^{n_b} z_j\Big) P_1\Big] 
+ Re\Big[\Big(\sum_{j=1}^{n_b} z_j^2\Big) P_2\Big] \label{beref2}\ee
where $z_j$ is the element of $Z_3$ which fixes the boundary conditions of
the $j$-th flavour. The property $z_j^3 = 1$ implies that the maximum power 
of $z_j$ that can appear in eq.\ \ref{beref2} is two.
The expression above should be compared with the corresponding fermion case 
with $n_f$ flavours and ordinary antiperiodic boundary conditions: 
\be S_{eff} = S_G - \frac{n_f}{2} [Re(P_0) + Re(P_1) + Re(P_2)] \ee

The bermion boundary conditions can change the signs of $P_1$ and $P_2$ terms 
in the effective action: for example with two bermion flavours and the choice
\[ z_1 = \exp(+2\pi i/3) \]
\[ z_2 = \exp(-2\pi i/3) \]
one gets:
\be S_{eff} = S_G +2 Re(P_0) - Re(P_1) - Re(P_2) \ee
i.e. the contribution of two bermion flavours has been changed,
 for the second class of diagrams, into the one of two fermion flavours.
We introduce a ``normal'' fermion number $n_f$ and two ``magnetic'' 
ones, $m_{f1}$ and $m_{f2}$ to distinguish between loops with winding 
number $1$ and $2$ respectively and we write the effective action in the form: 
\be S_{eff} = S_G - \frac{n_f}{2} Re(P_0) - \frac{m_{f1}}{2} Re(P_1) - 
\frac{m_{f2}}{2} Re(P_2) \ee

In table 1 we give, for various numbers of bermion flavours,
the values of $n_f$, $m_{f1}$ and $m_{f2}$ that are obtained with
specific choices of the boundary conditions. 
The correct fermion magnetic flavour numbers can be obtained only in
some special cases. The extrapolations to positive $n_f$ at fixed
values of $m_{f1}$ and $m_{f2}$ can be made only for a few values: 
for example, for $m_{f1} = m_{f2} = 2$ the available $n_f$ are $-5$ and  $-2$.
For the other cases the table includes a further choice of boundary 
conditions: the periodic ones ($z_j = -1$). By adopting these conditions for 
some of the bermion flavours we can simulate theories where the values of 
$m_{f1}$ and $m_{f2}$ are different and also the contributions of terms with 
winding number larger than 2 are modified. In the case of moderately heavy 
quark masses where winding numbers higher than 2 can be neglected, the 
$m_{f2}$ contributions can be seen as a small correction to the $m_{f1}$ 
terms. In this case thay can be extrapolated together with $n_f$ terms. 
In the heavy quark mass case also the terms with winding numbers 2 can be 
neglected and the fate of $m_{f2}$ terms ignored. 

In the coming section we will investigate the role of the winding number 
2 term: at $N_T=4$ and for $\kappa > 0.14$ we find some evidence 
for the relevance of such a term.

\section{Comparison with existing results} 

We have tested our method at $N_t=2$ with the results of the 
hopping parameter expansion. According to ref. \cite{karsch}, the
effective action in this approximation and at leading order reads:
\be S_{eff} = S_G + H\sum_{\vec{x}}Re\, Tr \prod_{t=1}^{N_t}U_{\hat 0}(\vec{x},t) 
+ o(\kappa^4) \label{hpe}\ee
where $H = 2m_{f1}(2\kappa)^{N_t}$. The shift of the critical $\beta$ value 
$\beta_c(H)$ from $\beta_c(0)$ can be predicted \cite{karsch} in term of the 
jumps of Polyakov loop and plaquette expectation values in the pure 
gauge theory:
\be \beta_c(H) = \beta_c(0) - (4.94\pm 0.75) H \label{bjump}\ee

To leeding order in $\kappa$ only terms with winding number 1 contribute 
and only the value of $m_{f1}$ is relevant. Periodic boundary 
conditions instead of antiperiodic ones are sufficient to promote bermion 
effects into fermion effects. 
In order to test the expression above, we have run 
different values of $n_f$ (from $-2$ to $-10$), $m_{f1}$ (from 2 to 6) 
and $\kappa$ (up to 0.071), while keeping the coefficient $8m_{f1}\kappa^2$ 
of the $Z_3$ breaking term in eq.\ \ref{hpe} equal to a fixed value ranging 
from 0.01 to 0.08 and we have measured Polyakov loops and plaquet\-tes. 
For this small values of $\kappa$ the effects of different values of $n_f$, 
which are of higher order in the hopping expansion, should be negligible 
and in fact the values of the observables at fixed $\beta$ depend 
only upon $8m_{f1}\kappa^2$. We have determined $\beta_c(H)$ by 
monitoring the jump in the Polyakov loop expectation value. The 
results, shown in fig.\ 1, are in remarkable agreement with the 
prediction of eq.\ \ref{bjump}, confirming the validity of the 
method in the hopping expansion regime. 

Increasing $N_t$ and going to higher values of $\kappa$ rapresent a more severe 
test for the method. We have compared our results for $N_t = 4$ 
with those of hybrid Monte Carlo simulations of ref.\ \cite{frabernardo}, and 
in particular we have studied the behaviour of the Polyakov loops 
and of the plaquettes around the phase transition point.
In this case the $m_{f1}$ term starts at the order $\kappa^4$, the 
renormalization effects due to ``normal'' fermion loops are
not negligible anymore and we have to perform an extrapolation from 
negative values, at fixed values of $m_{f1}$. As already discussed,
the values of $m_{f2}$ cannot be fixed at the fermionic ones at all values 
of $n_f$. In figures 2-4 we present the behaviour as a function of $\beta$ 
of the space-space plaquette 1-$P_\sigma = \frac{1}{N_c}Re\langle 
Tr (U_1U_2U^\dagger_3U^\dagger_4) \rangle$
and of the absolute value of the Polyakov loop $L$, defined by:
\be L = \frac{1}{N_s}\sum_{\vec{x}} \frac{1}{N_c} 
Tr\prod_{t=1}^{N_t}U_{\hat{0}}(\vec{x},t) \ee 
for different values of $n_f$ and constant values of $m_{f1}$ at 
$\kappa = 0.12$, 0.14 and 0.16 respectively, together with the 
extrapolations to full QCD with two fermions ($n_f = m_{f1} = m_{f2}=2$). 
In figures 5 we present for the case $\kappa = 0.14$ the results for the 
difference between space-space and space-time plaquettes.

The procedure adopted for the extrapolation is the following: 
we keep fixed the values of $\kappa$ and of a given observable 
($P_\sigma$, $|P_\sigma-P_\tau|$ or $|L|$) 
and extrapolate the corresponding $\beta$ as 
a function of $n_f$. From previous applications of the bermion method 
to QCD at zero temperature it is known that, in the intermediate quark mass 
region, a good extrapolation in $n_f$ is obtained by keeping 
fixed {\it renormalized} quantities. This is true in principle also for 
the simulations at finite temperature, however it must be noticed 
that our results refer, at least across 
the phase transition jump, to a rather heavy quark mass region where the 
renormalization effects are not so large to require a non perturbative 
treatment and therefore the extrapolation at fixed bare parameters gives 
a good first guess. 

The results refer to the cases $a$, $b$, $c$ and $e$ 
of table 1 which correspond to $m_{f1} = 2$ and four different negative 
values of $n_f$. For small values of $\kappa$ the effect of the terms 
with winding number larger than one should be negligible and 
therefore, indipendently from the value of $m_{f2}$ the data should align 
as a function of $n_f$, as is the case for $\kappa = 0.12$ and 
0.14. Even if they do not align we can consider two possible 
extrapolations which take into account the relevance of $m_{f2}$ terms.
The first one is from the cases $b$ and $e$ where all the $Z_3$ breaking 
terms are fixed at the fermion value $m_{f1} = m_{f2} = 2$. The other one 
is to extrapolate from cases $a$ and $c$: it keeps fixed 
at the fermion value only the dominant $Z_3$ breaking term ($m_{f1} = 2$), 
while $m_{f2}$ is extrapolated from negative to positive values together 
with $n_f$. The results of the two extrapolations are given in the figures: 
they agree with each other and with the 
results of direct hybrid Monte Carlo simulations 
\cite{frabernardo} showing the reliability of the method beyond 
the hopping expansion of eq.\ \ref{hpe}.

At $\kappa = 0.16$ we only show the full results for $n_b = 1,2,3$. 
The case $n_b = 2$ shows a different shape with respect to the other 
two and indicates the relevance of $m_{f2}$ terms. The 
simulations with $n_b = 5$ are only shown for the lowest values 
of $\beta$. We have observed that in this case the correlation times 
increase by almost two order of magnitude showing a severe critical 
slowing down. This may be ascribed to the delicate cancellation of 
``magnetic'' effects among three out the five bermion flavours. 
We plan to further study algorithms to improve the thermalization in 
a multiboson environment. In the region where we can perform the separate 
extrapolations of $n_b = 1, 3$ and $n_b = 2, 5$ results we observe again 
a reasonable consistency among them.

We have shown that, by suitably modifying the boundary conditions,
the bermion method can be applied to the study of full QCD at finite
temperature and compared successfully with existing results.
A detailed analysis of the nature of the phase transition, of the form of
the effective action and of the possible interruption
of the phase transition line for lower values of quark masses is currently 
under study \cite{Noi8}.

\vskip 0.8 truecm 
{\bf Acknowledgments.} 
We thank F.\ Karsch for many stimulating discussions.

\newpage

\begin{table}
\begin{center}
\begin{tabular}{|| c | c | l | c | c | c ||}\hline 
case & $n_b$ & $z_j$                          & $n_f$ & $m_{f1}$ & $m_{f2}$ \\ 
\hline 
a &   1   & $z_1$ = $-1$                   & $-$2  &     2    &   $-$2   \\ 
\hline 
b &   2   & $z_1$ = $e^{+\frac{2\pi}{3}i}$ & $-$4  &     2    &     2    \\ 
&       & $z_2$ = $e^{-\frac{2\pi}{3}i}$ & & & \\
\hline 
c &   3   & $z_1$ = $z_2$ = $-1$           & $-$6  &     2    &   $-$6   \\ 
&       & $z_3$ = 1                      & & & \\
\hline
&       & $z_1$ = $e^{+\frac{2\pi}{3}i}$ & & & \\
d &   4   & $z_2$ = $e^{-\frac{2\pi}{3}i}$ & $-$8 &      2    &   $-$2   \\ 
&       & $z_3$ = 1                              & & & \\
&       & $z_4$ = $-1$                             & & & \\
\hline
&       & $z_1$ = $z_2$ = $e^{+\frac{2\pi}{3}i}$ & & & \\
e &   5   & $z_3$ = $z_4$ = $e^{-\frac{2\pi}{3}i}$ & $-$10 &     2    &     2   \\ 
&       & $z_5$ = 1                              & & & \\
\hline 
\hline 
f &   2   & $z_1$ = $-1$                   & $-$4  &     0    &   $-$4   \\ 
&       & $z_2$ = 1                      & & & \\
\hline 
&       & $z_1$ = $e^{+\frac{2\pi}{3}i}$ & & & \\
g &   3   & $z_2$ = $e^{-\frac{2\pi}{3}i}$ & $-$6 &     0    &     0     \\ 
&       & $z_3$ = 1                      & & & \\
\hline 
h &   4   & $z_1$ = $z_2$ = $-1$           & $-$8  &     0    &   $-$8   \\ 
&       & $z_3$ = $z_4$ = 1              & & & \\
\hline 
&       & $z_1$ = $e^{+\frac{2\pi}{3}i}$ & & & \\
i &   5   & $z_2$ = $e^{-\frac{2\pi}{3}i}$ & $-$10 &     0    &   $-$4   \\ 
&       & $z_3$ = $z_4$ = 1              & & & \\
&       & $z_5$ = $-1$                   & & & \\
\hline
\hline
l &  $n$  & $z_j$ = $-1$ ($j$=1,...,$n$)   & $-2n$ &   $2n$   &   $-2n$  \\
\hline
\end{tabular}
\end{center}
\caption{The effective flavour numbers $n_f$, $m_{f1}$ and $m_{f2}$ obtained 
with boundary conditions $z_j$ in theories with $n_b$ bermion fields.}
\end{table}

\begin{figure} 
\begin{center}
\setlength{\unitlength}{0.240900pt}
\ifx\plotpoint\undefined\newsavebox{\plotpoint}\fi
\sbox{\plotpoint}{\rule[-0.175pt]{0.350pt}{0.350pt}}%
\begin{picture}(1574,1800)(0,0)
\tenrm
\sbox{\plotpoint}{\rule[-0.175pt]{0.350pt}{0.350pt}}%
\put(264,158){\rule[-0.175pt]{4.818pt}{0.350pt}}
\put(242,158){\makebox(0,0)[r]{4.5}}
\put(1490,158){\rule[-0.175pt]{4.818pt}{0.350pt}}
\put(264,376){\rule[-0.175pt]{4.818pt}{0.350pt}}
\put(242,376){\makebox(0,0)[r]{4.6}}
\put(1490,376){\rule[-0.175pt]{4.818pt}{0.350pt}}
\put(264,595){\rule[-0.175pt]{4.818pt}{0.350pt}}
\put(242,595){\makebox(0,0)[r]{4.7}}
\put(1490,595){\rule[-0.175pt]{4.818pt}{0.350pt}}
\put(264,813){\rule[-0.175pt]{4.818pt}{0.350pt}}
\put(242,813){\makebox(0,0)[r]{4.8}}
\put(1490,813){\rule[-0.175pt]{4.818pt}{0.350pt}}
\put(264,1032){\rule[-0.175pt]{4.818pt}{0.350pt}}
\put(242,1032){\makebox(0,0)[r]{4.9}}
\put(1490,1032){\rule[-0.175pt]{4.818pt}{0.350pt}}
\put(264,1250){\rule[-0.175pt]{4.818pt}{0.350pt}}
\put(242,1250){\makebox(0,0)[r]{5}}
\put(1490,1250){\rule[-0.175pt]{4.818pt}{0.350pt}}
\put(264,1469){\rule[-0.175pt]{4.818pt}{0.350pt}}
\put(242,1469){\makebox(0,0)[r]{5.1}}
\put(1490,1469){\rule[-0.175pt]{4.818pt}{0.350pt}}
\put(264,1687){\rule[-0.175pt]{4.818pt}{0.350pt}}
\put(242,1687){\makebox(0,0)[r]{5.2}}
\put(1490,1687){\rule[-0.175pt]{4.818pt}{0.350pt}}
\put(264,158){\rule[-0.175pt]{0.350pt}{4.818pt}}
\put(264,113){\makebox(0,0){0}}
\put(264,1667){\rule[-0.175pt]{0.350pt}{4.818pt}}
\put(389,158){\rule[-0.175pt]{0.350pt}{4.818pt}}
\put(389,113){\makebox(0,0){0.01}}
\put(389,1667){\rule[-0.175pt]{0.350pt}{4.818pt}}
\put(513,158){\rule[-0.175pt]{0.350pt}{4.818pt}}
\put(513,113){\makebox(0,0){0.02}}
\put(513,1667){\rule[-0.175pt]{0.350pt}{4.818pt}}
\put(638,158){\rule[-0.175pt]{0.350pt}{4.818pt}}
\put(638,113){\makebox(0,0){0.03}}
\put(638,1667){\rule[-0.175pt]{0.350pt}{4.818pt}}
\put(762,158){\rule[-0.175pt]{0.350pt}{4.818pt}}
\put(762,113){\makebox(0,0){0.04}}
\put(762,1667){\rule[-0.175pt]{0.350pt}{4.818pt}}
\put(887,158){\rule[-0.175pt]{0.350pt}{4.818pt}}
\put(887,113){\makebox(0,0){0.05}}
\put(887,1667){\rule[-0.175pt]{0.350pt}{4.818pt}}
\put(1012,158){\rule[-0.175pt]{0.350pt}{4.818pt}}
\put(1012,113){\makebox(0,0){0.06}}
\put(1012,1667){\rule[-0.175pt]{0.350pt}{4.818pt}}
\put(1136,158){\rule[-0.175pt]{0.350pt}{4.818pt}}
\put(1136,113){\makebox(0,0){0.07}}
\put(1136,1667){\rule[-0.175pt]{0.350pt}{4.818pt}}
\put(1261,158){\rule[-0.175pt]{0.350pt}{4.818pt}}
\put(1261,113){\makebox(0,0){0.08}}
\put(1261,1667){\rule[-0.175pt]{0.350pt}{4.818pt}}
\put(1385,158){\rule[-0.175pt]{0.350pt}{4.818pt}}
\put(1385,113){\makebox(0,0){0.09}}
\put(1385,1667){\rule[-0.175pt]{0.350pt}{4.818pt}}
\put(1510,158){\rule[-0.175pt]{0.350pt}{4.818pt}}
\put(1510,113){\makebox(0,0){0.1}}
\put(1510,1667){\rule[-0.175pt]{0.350pt}{4.818pt}}
\put(264,158){\rule[-0.175pt]{300.161pt}{0.350pt}}
\put(1510,158){\rule[-0.175pt]{0.350pt}{368.336pt}}
\put(264,1687){\rule[-0.175pt]{300.161pt}{0.350pt}}
\put(45,922){\makebox(0,0)[l]{\shortstack{{\large $\beta_c$}}}}
\put(887,23){\makebox(0,0){{\large $8m_{f1}k^2$}}}
\put(264,158){\rule[-0.175pt]{0.350pt}{368.336pt}}
\put(264,1450){\makebox(0,0){$\star$}}
\put(389,1346){\makebox(0,0){$\star$}}
\put(762,999){\makebox(0,0){$\star$}}
\put(1261,486){\makebox(0,0){$\star$}}
\put(264,1449){\rule[-0.175pt]{0.350pt}{0.482pt}}
\put(254,1449){\rule[-0.175pt]{4.818pt}{0.350pt}}
\put(254,1451){\rule[-0.175pt]{4.818pt}{0.350pt}}
\put(389,1344){\rule[-0.175pt]{0.350pt}{0.964pt}}
\put(379,1344){\rule[-0.175pt]{4.818pt}{0.350pt}}
\put(379,1348){\rule[-0.175pt]{4.818pt}{0.350pt}}
\put(762,966){\rule[-0.175pt]{0.350pt}{15.899pt}}
\put(752,966){\rule[-0.175pt]{4.818pt}{0.350pt}}
\put(752,1032){\rule[-0.175pt]{4.818pt}{0.350pt}}
\put(1261,376){\rule[-0.175pt]{0.350pt}{52.757pt}}
\put(1251,376){\rule[-0.175pt]{4.818pt}{0.350pt}}
\put(1251,595){\rule[-0.175pt]{4.818pt}{0.350pt}}
\sbox{\plotpoint}{\rule[-0.250pt]{0.500pt}{0.500pt}}%
\put(264,1450){\usebox{\plotpoint}}
\put(264,1450){\usebox{\plotpoint}}
\put(278,1435){\usebox{\plotpoint}}
\put(293,1420){\usebox{\plotpoint}}
\put(308,1406){\usebox{\plotpoint}}
\put(322,1391){\usebox{\plotpoint}}
\put(337,1376){\usebox{\plotpoint}}
\put(352,1362){\usebox{\plotpoint}}
\put(366,1347){\usebox{\plotpoint}}
\put(381,1332){\usebox{\plotpoint}}
\put(396,1318){\usebox{\plotpoint}}
\put(410,1303){\usebox{\plotpoint}}
\put(425,1288){\usebox{\plotpoint}}
\put(440,1274){\usebox{\plotpoint}}
\put(455,1259){\usebox{\plotpoint}}
\put(469,1244){\usebox{\plotpoint}}
\put(484,1230){\usebox{\plotpoint}}
\put(499,1215){\usebox{\plotpoint}}
\put(513,1200){\usebox{\plotpoint}}
\put(528,1186){\usebox{\plotpoint}}
\put(543,1171){\usebox{\plotpoint}}
\put(557,1156){\usebox{\plotpoint}}
\put(572,1142){\usebox{\plotpoint}}
\put(587,1127){\usebox{\plotpoint}}
\put(601,1112){\usebox{\plotpoint}}
\put(616,1098){\usebox{\plotpoint}}
\put(631,1083){\usebox{\plotpoint}}
\put(646,1068){\usebox{\plotpoint}}
\put(660,1054){\usebox{\plotpoint}}
\put(675,1039){\usebox{\plotpoint}}
\put(690,1024){\usebox{\plotpoint}}
\put(704,1010){\usebox{\plotpoint}}
\put(719,995){\usebox{\plotpoint}}
\put(734,980){\usebox{\plotpoint}}
\put(748,966){\usebox{\plotpoint}}
\put(763,951){\usebox{\plotpoint}}
\put(778,936){\usebox{\plotpoint}}
\put(792,922){\usebox{\plotpoint}}
\put(807,907){\usebox{\plotpoint}}
\put(822,892){\usebox{\plotpoint}}
\put(837,878){\usebox{\plotpoint}}
\put(851,863){\usebox{\plotpoint}}
\put(866,848){\usebox{\plotpoint}}
\put(881,834){\usebox{\plotpoint}}
\put(895,819){\usebox{\plotpoint}}
\put(910,805){\usebox{\plotpoint}}
\put(925,790){\usebox{\plotpoint}}
\put(939,775){\usebox{\plotpoint}}
\put(954,761){\usebox{\plotpoint}}
\put(969,746){\usebox{\plotpoint}}
\put(984,731){\usebox{\plotpoint}}
\put(998,717){\usebox{\plotpoint}}
\put(1013,702){\usebox{\plotpoint}}
\put(1028,687){\usebox{\plotpoint}}
\put(1042,673){\usebox{\plotpoint}}
\put(1057,658){\usebox{\plotpoint}}
\put(1072,643){\usebox{\plotpoint}}
\put(1086,629){\usebox{\plotpoint}}
\put(1101,614){\usebox{\plotpoint}}
\put(1116,599){\usebox{\plotpoint}}
\put(1130,585){\usebox{\plotpoint}}
\put(1145,570){\usebox{\plotpoint}}
\put(1160,555){\usebox{\plotpoint}}
\put(1175,541){\usebox{\plotpoint}}
\put(1189,526){\usebox{\plotpoint}}
\put(1204,511){\usebox{\plotpoint}}
\put(1219,497){\usebox{\plotpoint}}
\put(1233,482){\usebox{\plotpoint}}
\put(1248,467){\usebox{\plotpoint}}
\put(1263,453){\usebox{\plotpoint}}
\put(1277,438){\usebox{\plotpoint}}
\put(1292,423){\usebox{\plotpoint}}
\put(1307,409){\usebox{\plotpoint}}
\put(1321,394){\usebox{\plotpoint}}
\put(1336,379){\usebox{\plotpoint}}
\put(1351,365){\usebox{\plotpoint}}
\put(1366,350){\usebox{\plotpoint}}
\put(1380,335){\usebox{\plotpoint}}
\put(1395,321){\usebox{\plotpoint}}
\put(1410,306){\usebox{\plotpoint}}
\put(1424,291){\usebox{\plotpoint}}
\put(1439,277){\usebox{\plotpoint}}
\put(1454,262){\usebox{\plotpoint}}
\put(1468,247){\usebox{\plotpoint}}
\put(1483,233){\usebox{\plotpoint}}
\put(1498,218){\usebox{\plotpoint}}
\put(1510,207){\usebox{\plotpoint}}
\put(264,1450){\usebox{\plotpoint}}
\put(264,1450){\usebox{\plotpoint}}
\put(280,1437){\usebox{\plotpoint}}
\put(297,1425){\usebox{\plotpoint}}
\put(314,1413){\usebox{\plotpoint}}
\put(330,1400){\usebox{\plotpoint}}
\put(347,1388){\usebox{\plotpoint}}
\put(364,1376){\usebox{\plotpoint}}
\put(381,1364){\usebox{\plotpoint}}
\put(397,1351){\usebox{\plotpoint}}
\put(414,1339){\usebox{\plotpoint}}
\put(431,1327){\usebox{\plotpoint}}
\put(448,1314){\usebox{\plotpoint}}
\put(464,1302){\usebox{\plotpoint}}
\put(481,1290){\usebox{\plotpoint}}
\put(498,1278){\usebox{\plotpoint}}
\put(514,1265){\usebox{\plotpoint}}
\put(531,1253){\usebox{\plotpoint}}
\put(548,1241){\usebox{\plotpoint}}
\put(565,1228){\usebox{\plotpoint}}
\put(581,1216){\usebox{\plotpoint}}
\put(598,1204){\usebox{\plotpoint}}
\put(615,1192){\usebox{\plotpoint}}
\put(632,1179){\usebox{\plotpoint}}
\put(648,1167){\usebox{\plotpoint}}
\put(665,1155){\usebox{\plotpoint}}
\put(682,1142){\usebox{\plotpoint}}
\put(698,1130){\usebox{\plotpoint}}
\put(715,1118){\usebox{\plotpoint}}
\put(732,1106){\usebox{\plotpoint}}
\put(749,1093){\usebox{\plotpoint}}
\put(765,1081){\usebox{\plotpoint}}
\put(782,1069){\usebox{\plotpoint}}
\put(799,1056){\usebox{\plotpoint}}
\put(816,1044){\usebox{\plotpoint}}
\put(832,1032){\usebox{\plotpoint}}
\put(849,1020){\usebox{\plotpoint}}
\put(866,1007){\usebox{\plotpoint}}
\put(882,995){\usebox{\plotpoint}}
\put(899,983){\usebox{\plotpoint}}
\put(916,970){\usebox{\plotpoint}}
\put(933,958){\usebox{\plotpoint}}
\put(949,946){\usebox{\plotpoint}}
\put(966,934){\usebox{\plotpoint}}
\put(983,921){\usebox{\plotpoint}}
\put(1000,909){\usebox{\plotpoint}}
\put(1016,897){\usebox{\plotpoint}}
\put(1033,884){\usebox{\plotpoint}}
\put(1050,872){\usebox{\plotpoint}}
\put(1067,860){\usebox{\plotpoint}}
\put(1083,848){\usebox{\plotpoint}}
\put(1100,835){\usebox{\plotpoint}}
\put(1117,823){\usebox{\plotpoint}}
\put(1133,811){\usebox{\plotpoint}}
\put(1150,798){\usebox{\plotpoint}}
\put(1167,786){\usebox{\plotpoint}}
\put(1184,774){\usebox{\plotpoint}}
\put(1200,762){\usebox{\plotpoint}}
\put(1217,749){\usebox{\plotpoint}}
\put(1234,737){\usebox{\plotpoint}}
\put(1251,725){\usebox{\plotpoint}}
\put(1267,712){\usebox{\plotpoint}}
\put(1284,700){\usebox{\plotpoint}}
\put(1301,688){\usebox{\plotpoint}}
\put(1317,676){\usebox{\plotpoint}}
\put(1334,663){\usebox{\plotpoint}}
\put(1351,651){\usebox{\plotpoint}}
\put(1368,639){\usebox{\plotpoint}}
\put(1384,626){\usebox{\plotpoint}}
\put(1401,614){\usebox{\plotpoint}}
\put(1418,602){\usebox{\plotpoint}}
\put(1435,590){\usebox{\plotpoint}}
\put(1451,577){\usebox{\plotpoint}}
\put(1468,565){\usebox{\plotpoint}}
\put(1485,553){\usebox{\plotpoint}}
\put(1501,540){\usebox{\plotpoint}}
\put(1510,535){\usebox{\plotpoint}}
\end{picture}
\caption{The critical value of $\beta$ in bermion simulations on a 
$16^3\times 2$ lattice as a function of $8m_{f1}\kappa^2$ is compared 
with the hopping expansion prediction [2] (see eq.\ 11).}
\end{center}
\end{figure}
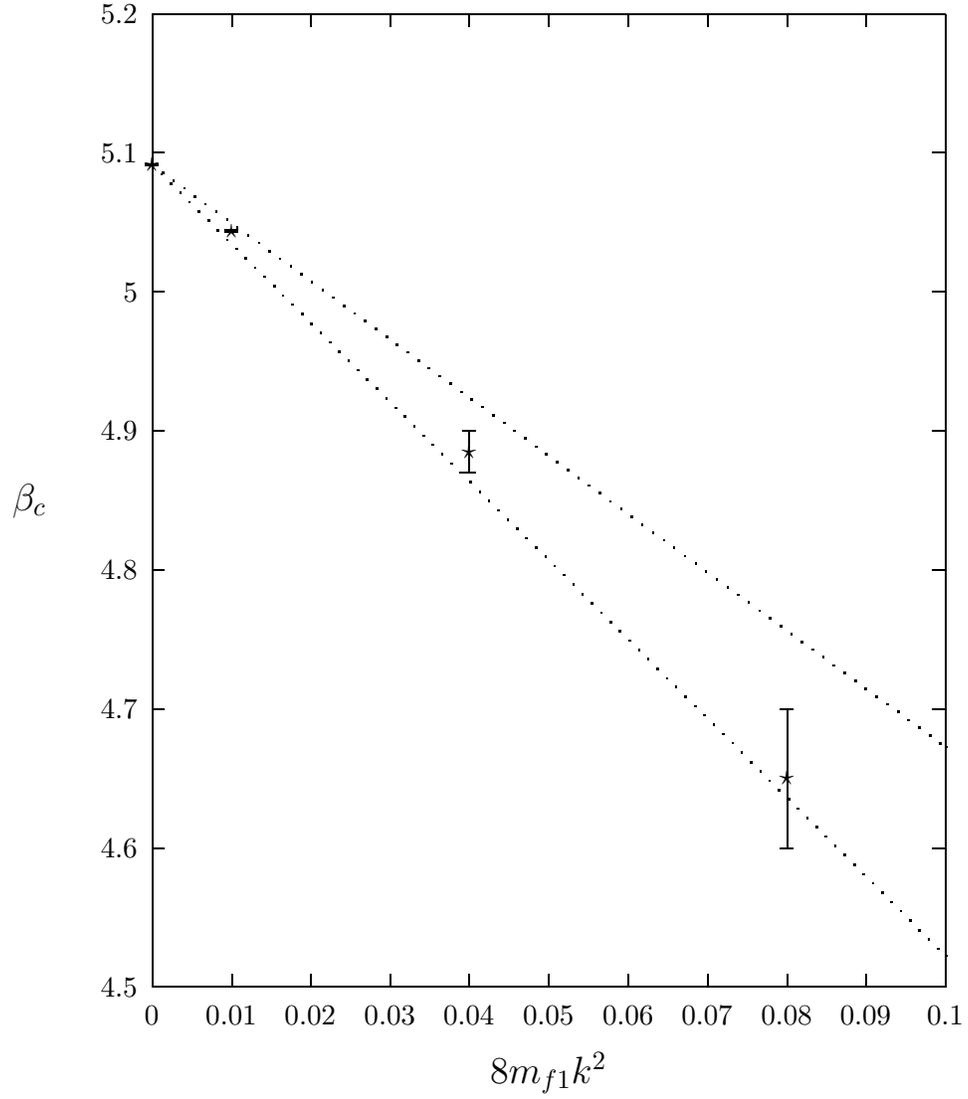

\newpage

\begin{figure} 
\begin{center}
\input{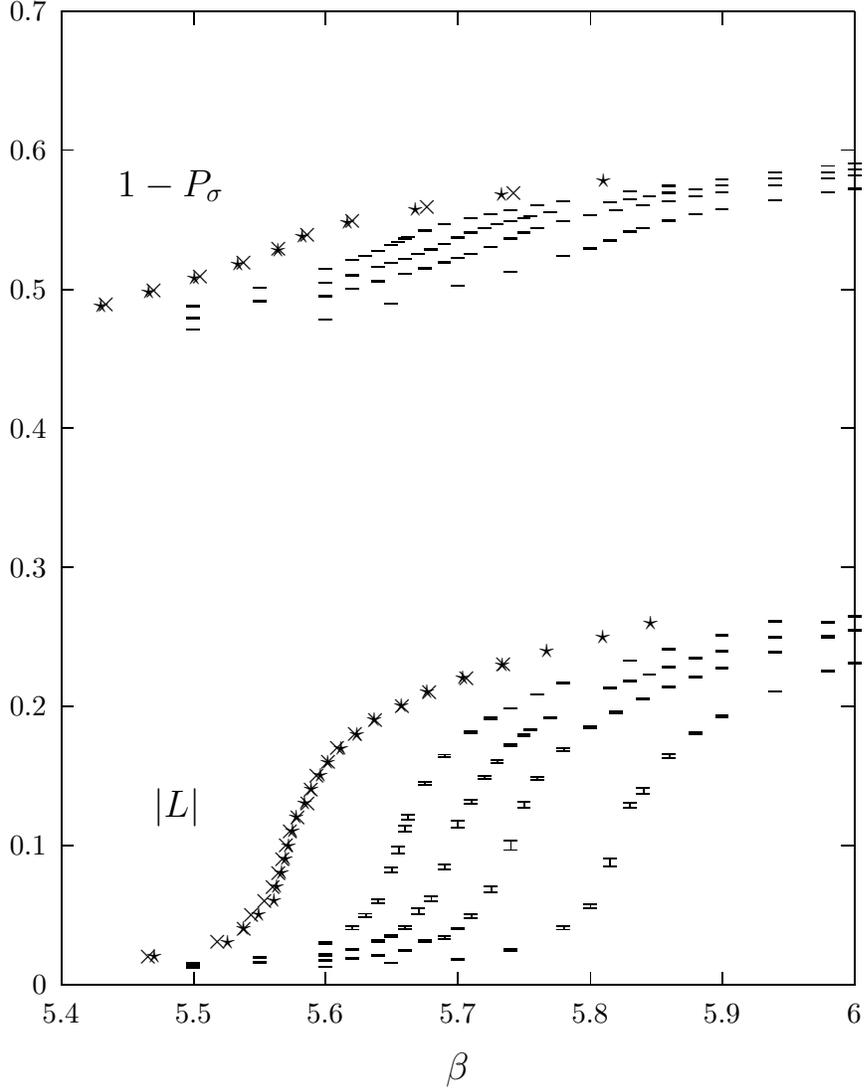}
\caption{The average space-space plaquette $1-P_\sigma$ and the absolute 
value of 
the Polyakov loop $|L|$ on a $16^3\times 4$ lattice for $\kappa = 0.12$ 
for four simulations with negative $n_f$ (corresponding, from right to left, 
to cases $e$, $c$, $b$, $a$ of table 1) and the extrapolations to 
the fermion case $n_f=2$ discussed in the text ($\star$: extrapolation from 
cases $a$ and $c$; $\times$: extrapolation from cases $b$ and $e$). }
\end{center}
\end{figure}

\newpage

\begin{figure} 
\begin{center}
\input{fig3}
\caption{The same as in fig.\ 2 for $\kappa = 0.14$.}
\end{center}
\end{figure}

\newpage

\begin{figure} 
\begin{center}
\input{fig4}
\caption{The same as in fig.\ 2 for $\kappa = 0.16$.}
\end{center}
\end{figure}


\newpage
\begin{figure} 
\begin{center}
\input{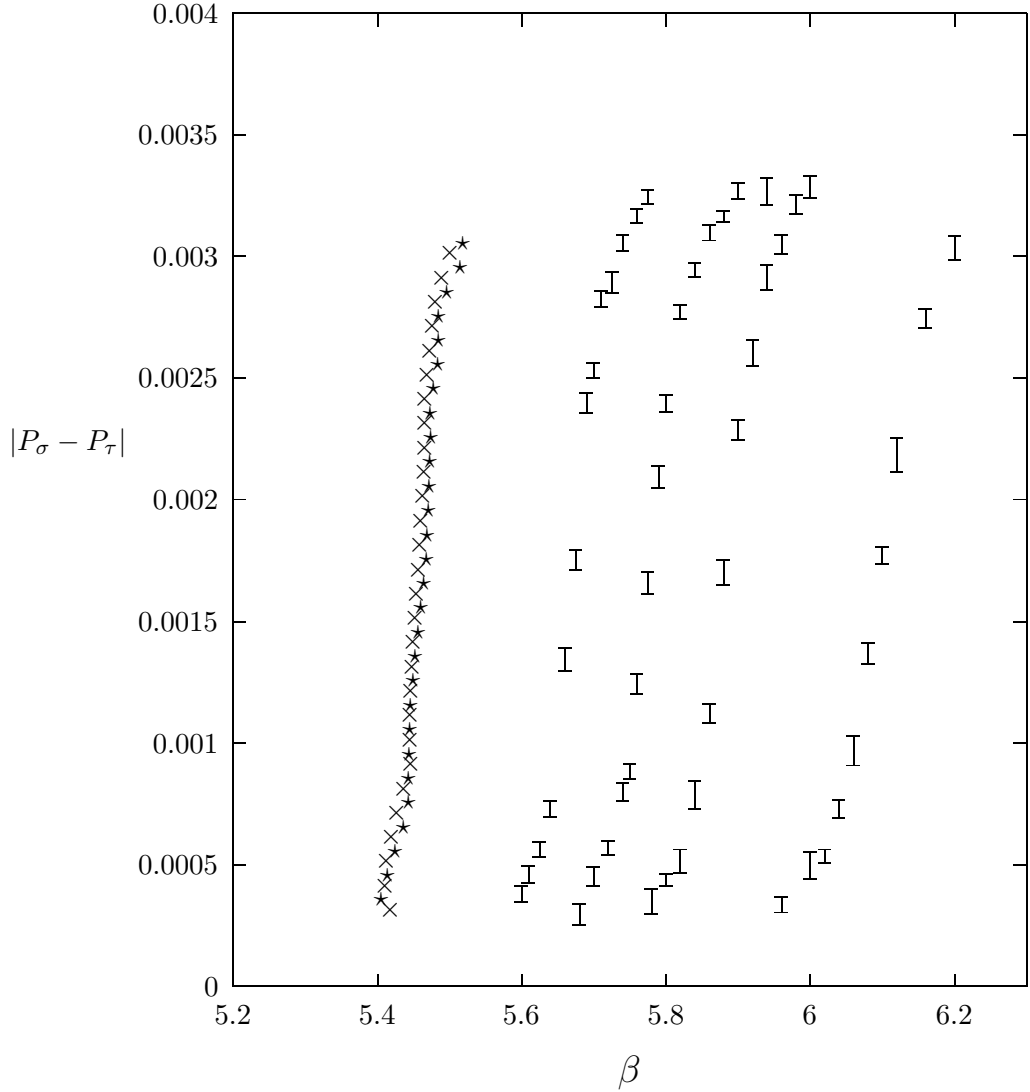}
\caption{The difference between space-space and space-time plaquettes 
for $\kappa = 0.14$ for four simulations with negative $n_f$ 
(corresponding, from right to left, to cases $e$, $c$, $b$, $a$ of table 1) 
and the extrapolations to the fermion case $n_f=2$ discussed in the text 
($\star$: extrapolation from cases $a$ and $c$; $\times$: extrapolation from 
cases $b$ and $e$).}
\end{center}
\end{figure}



\begin{thebibliography}{99}

\bibitem{LL}
S.\ J.\ Anthony, C.\ H.\ Llewellyn Smith and J.\ F.\ Wheater, 
Phys.\ Lett.\ B 116 (1982) 287.
 
\bibitem{Noi} 
R.\ Petronzio, Nucl.\ Phys.\ B (Proc.\ Suppl.) 42 (1995) 942; 
G.\ M.\ de Divitiis, R.\ Frezzotti, M.\ Guagnelli, 
M.\ Masetti, R.\ Petronzio, Nucl.\ Phys.\ B 455 (1995) 274; 
G.\ M.\ de Divitiis, R.\ Frezzotti, M.\ Guagnelli, 
M.\ Masetti, R.\ Petronzio, Ph.\ Lett.\ B 367 (1996) 279; 
G.\ M.\ de Divitiis, R.\ Frezzotti, 
M.\ Masetti, R.\ Petronzio, preprints 
ROM2F-96-10, hep-lat/9603020 and ROM2F-96-16, hep/lat/9605002, 
submitted to Ph.\ Lett. B; 
R.\ Frezzotti, M.\ Masetti, R.\ Petronzio, preprint ROM2F-96-28 hep-lat/9605044 
submitted to Nucl.\ Phys.\ B. 

\bibitem{karsch} 
P.\ Hasenfraz, F.\ Karsch, I.\ O.\ Stamatescu, Phys.\ Lett.\ B 133 (1983) 221; 
F.\ Green, F.\ Karsch, Nucl.\ Phys.\ B 238 (1984) 297.

\bibitem{frabernardo} 
C.\ Bernard et al., Nucl.\ Phys.\ B (Proc.\ Suppl.) 34 (1994) 324; Phys.\ Rev.\ 
D 49 (1994) 3574. 

\bibitem{Noi8} 
G.\ M.\ de Divitiis, M.\ Masetti, R.\ Petronzio, in preparation. 


\end{thebibliography}
\end{document}